\documentclass[aps,prb,twocolumn,amsmath,amssymb,superscriptaddress]{revtex4}
\usepackage{graphicx}
\usepackage{bm}
\DeclareMathOperator{\Imag}{Im}
\begin{document}
\def \brho{{\hbox{\boldmath $\rho$}}}
\def \beps{{\hbox{\boldmath $\epsilon$}}}
\def \bdelta{{\hbox{\boldmath $\delta$}}}
\title{Core level spectra in XPS of pristine and doped graphene}

\author{V. Despoja}
\email{vito@phy.hr}
\affiliation{Department of Physics, University of Zagreb, Bijeni\v{c}ka 32, HR-10000 Zagreb, Croatia}
\affiliation{Donostia International Physics Center (DIPC), Paseo de Manuel de Lardizabal 4, ES-20018 San Sebastian, Spain}
\affiliation{Centro de Fisica de Materiales CSIC-UPV/EHU-MPC, Paseo de Manuel de Lardizabal 5, ES-20018 San Sebastian, Spain}
\author{M \v Sunji\' c}
\email{msunjic@phy.hr}
\affiliation{Department of Physics, University of Zagreb, Bijeni\v{c}ka 32, HR-10000 Zagreb, Croatia}
\affiliation{Donostia International Physics Center (DIPC), Paseo de Manuel de Lardizabal 4, ES-20018 San Sebastian, Spain}

\begin{abstract}
Spectra of the C1s core hole, created in XPS and screened by electronic excitations in 
pristine and doped graphene, are calculated and discussed. We find that singular effects in the 
lineshapes are not possible in the pristine graphene, and their observation should be connected 
with the doping. However, the structure of the low energy excitation spectrum in the region 
where the singular behaviour is expected leads to asymmetries in the core hole lineshapes in pristine graphene 
similar to those in doped graphene. This makes the analysis more complex than in the case of metals 
and may lead to incorrect or incomplete interpretation of experimental results.    
\end{abstract}

\maketitle

\section{Introduction}
Excitation of localized levels in solids, especially in metals, can lead to a variety of many body 
scattering processes, as can be revealed e.g. in X-ray photoemission spectra from these 
levels. These phenomena and the information that they provided about the structure and 
local dynamics of electrons in such systems was among reasons that 
they were extensively studied even since the pioneering theoretical 
\cite{Mahan1,Nozieres,DSlineshape,Anderson,Langreth,Combescot,Mahan2,Minh1,Minh2,Schotte} 
and experimental \cite{citrin1,citrin2} studies. 
Observation of asymmetric lineshapes was often  used as an indicator of the metallicity of the system. 
The extension of these studies to new materials, like 
graphene, led to renewed interest in the application and interpretation of 
various spectroscopies to these systems.        
In this paper we therefore provide theoretical prediction of core level photoemission spectra from 
pristine and doped graphene, discuss in detail the core level lineshapes and the possibility 
that they reveal singular behaviour, as is the case in metals.   

In Sec.\ref{sec2} we briefly derive expressions for a localised level spectrum, corresponding to the 
$1s$ level of carbon, interacting  with electronic excitations, which are described by the nonlocal dynamically 
screened Coulomb interaction. 
We shall neglect extrinsic (photoelectron) scattering processes, in which case the measured 
photoelectron spectrum corresponds to the core hole spectrum. 
In Sec.\ref{sec3} we present an analytic discussion of possible singularities in XPS 
lineshapes using asymptotic expressions for the singularity index.                 
and conclude that they can occur only in doped graphene.  
This may in principle open the possibility to directly connect doping and line 
asymmetry, but later in a more exact calculation we show the validity of these 
asymptotic results. 
In Sec.\ref{sec4} we present methodology we use in the calculation of the propagator of dynamically screened 
Coulomb interaction in graphene. With this full propagator in Sec.\ref{sec5} we calculate singularity indices in pristine and 
doped graphene and compare them with the 
previous approximate results. 
We show that, unlike the situation in the metals, the singularity indices $\alpha$ vary in a much 
more complicated way. For $\omega=0$ they indeed start from their asymptotic values, 
but soon the reverse situation occurs: in pristine graphene $\alpha$ increases above the 
value in doped graphene, which decreases until it reaches the 2D plasmon peak. The 
fact is that $E_F$, hole decay linewidth, pair and plasmon energies all lie in the same energy region, so that 
detailed calculations become necessary.   	  
In Sec.\ref{sec6} we present a method for calculation of the complete core-hole 
spectrum and its various properties.   
In Sec.\ref{Allspe} we use this method to calculate and discuss the shapes and properties of 
C1s core hole spectra in graphene.  We also compare them with the existing experimental results, pointing out 
possible incorrect interpretation of measured 
C1s core hole lineshapes in XPS. 

\section{Derivation of the localised level spectrum}
\label{sec2}

\begin{figure}
\includegraphics[width=\columnwidth]{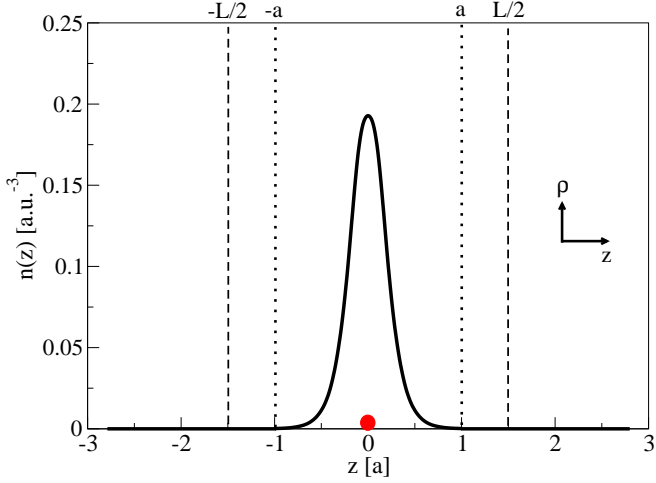}
\caption{LDA graphene electronic pseudodensity averaged over $xy$ plane. $L$ is the thickness of one supercell.
Scale on the abscissa is the unit cell parameter in $xy$ plane $a=4.651\ a.u.$. Red dot represents the core hole position.}
\label{Fig1}
\end{figure}

In Fig.\ref{Fig1} we show shematically the geometry of the system. Center of the quasi-two dimensional graphene 
layer is at the $z=0$ plane, and the core hole is created at some point ${\bf R}=(\brho,z)$, to be defined later, 
with the wave function $u({\bf r}-{\bf R})$. In this way our formalism can also be applied to study spectra of 
atoms adsorbed at various positions on graphene.               

The hamiltonian of the system is 
\[
H=H_0+H_{\textit{int}}
\]
where 
\begin{equation}
H_0=E_0d^+d+\sum_{{\bf K}n}E_{{\bf K}n}c^+_{{\bf K}n}c_{{\bf K}n},
\label{hamon_0}
\end{equation}
$E_0$ and $d^+$ are the energy and the creation operator of a C1s core state, 
and $C^+_{{\bf K}n}$ creates electrons in a graphene with energy 
$E_{{\bf K}n}$ and wave function $\Phi_{n{\bf K}}(\brho,z)$.
${\bf K}$ represents parallel wave vector and $n$ is the 
conduction/valence band quantum number. The interaction 
hamiltonian contains two terms   
\[
H_{\textit{int}}=V_1+V_2 
\]
where 
\begin{equation}
V_1=dd^+\sum_{ij}w_{i,j}c^+_{i}c_{j}
\label{hamon_1}
\end{equation}
represents interaction of core-hole with conduction/valence electrons in 
graphene and   
\begin{equation}
V_2=\sum_{ijkl}v_{ijkl}c^+_{i}c^+_{j}c_{k}c_{l}
\label{hamon_2}
\end{equation}
represents electron-electron interaction in graphene. In the last two formulas we introduced shorter notation for quantum numbers 
$i=({\bf K},n)$. The matrix elements are
\begin{eqnarray}
& w_{ij} = \int d{\bf r}\psi^*_{i}\left({\bf r}\right)U\left({\bf r,{\bf R}}\right)\psi_{j}\left({\bf r}\right) \nonumber \\
\nonumber \\
& v_{ijkl} = \int d{\bf r}d{\bf r}'\psi^*_{i}\left({\bf r}\right)\psi^*_{j}\left({\bf r}'\right)v\left({\bf r},{\bf r}'\right)
\psi_{k}\left({\bf r}'\right)\psi_{l}\left({\bf r}\right) \nonumber
\end{eqnarray}
with
\[
U\left({\bf r,{\bf R}}\right) = \int d{\bf r}'v\left({\bf r},{\bf r}'\right)\left|u\left({\bf r}'-{\bf R}\right)\right|^2 \nonumber \\
\]
The core-hole Green's function can be written as 
\[
G(t)=-i\theta(t)\ e^{-iE_0 t}\ \left\langle d\left|U(t,0)\right|d\right\rangle
\]
where $\left|d\right\rangle$ is a one hole state with the binding energy $E_0$ in the interacting Fermi 
sea of graphene electrons, and $U(t,0)$ is the evolution operator in the interaction representation, or 
as 
\[
G(t)=-i\theta(t)\ e^{-iE_0 t}\ e^{\Phi(t)}
\]
where $\Phi(t)$ is the sum of cumulants \cite{Singindex,Mahknj} 
\[
\Phi(t)=\phi(t)+\phi_d(t). 
\]      
$\phi_d(t)$ describes all processes responsible for the eventual core hole decay (Auger, radiative decay, etc.), and is 
usually given in the form 
\[
\phi_d(t)=e^{-\gamma|t|}
\] 
where $\gamma$ is the decay constant.   
$\phi(t)$ represents core-hole interaction with the graphene electrons, which leads to characteristic structures in the hole 
spectrum energy shift, satelite structures, etc.

\begin{figure}[h]
\includegraphics[width=5cm]{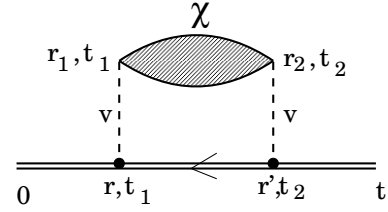}
\caption{The lowest order cumulant in the XPS process.}
\label{Fig2}
\end{figure}

\begin{figure}[h]
\centering
\includegraphics[width=2cm]{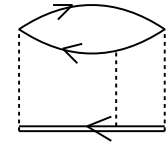}
\caption{Processes in which the hole interacts with the excited electron-hole pair.}
\label{Fig3}
\end{figure}

\begin{figure}[h]
\includegraphics[width=\columnwidth]{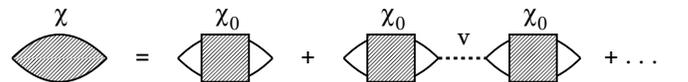}
\caption{Interacting electrons response function.}
\label{Fig4}
\end{figure}

\begin{figure}[h]
\includegraphics[width=3cm]{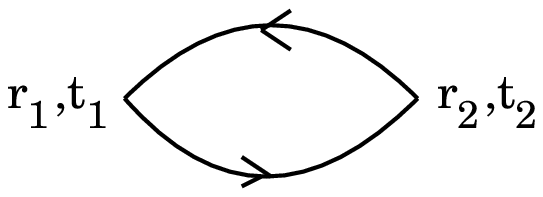}
\caption{Lowest order process in the expansion of $\chi_0$.}
\label{Fig5}
\end{figure}

In the following we shall make use of the formalism developed in  \cite{Singindex} to study XPS spectrum from localised 
levels in the vicinity of metallic surfaces following earlier work on bulk metals \cite{Mahan1,Nozieres,DSlineshape,Anderson,Langreth,Combescot,Mahan2,Minh1,Minh2,Schotte}. 
For the localised hole only the lowest 
order cumulant, shown in Fig.\ref{Fig2}, is finite. Here $v({\bf r}-{\bf r}')$ is the bare Coulomb interaction and $\chi$ is the exact response 
function of graphene electrons  which can be written as 
\[
\chi({\bf r}_1,{\bf r}_2,t_1,t_2)=\ \int^{\infty}_{-\infty}\frac{d\omega}{2\pi}e^{-i\omega t}\ \chi({\bf r}_1,{\bf r}_2,\omega) 
\]
In the expansion shown in Fig.\ref{Fig2} we neglect processes where the core hole directly couples to the excited electron-hole 
pair, as e.g. in Fig.\ref{Fig3}.   
Exact response function $\chi$ can be obtained as an infnite sum of diagrams shown in Fig.\ref{Fig4}.
Here $\chi_0$ describes excitation and annihilation of an electron-hole 
pair created by the $V_1$ potential, and in principle includes all their scattering processes due 
to the $V_2$ potential. In the spirit of RPA we shall take only the lowest order term, shown in 
Fig.\ref{Fig5}.  
From the diagram in Fig.\ref{Fig2} we obtain 
\begin{eqnarray}
\phi(t)=i \int^t_0dt_1\int^{t_1}_0dt_2\int d{\bf r}\int d{\bf r}'\int d{\bf r}_1\int d{\bf r}_2
\nonumber\\
\label{cuumulu}\\
|u({\bf r}-{\bf R})|^2v({\bf r}-{\bf r}_1)\chi({\bf r}_1,{\bf r}_2,t_1,t_2)v({\bf r}_2-{\bf r}')|u({\bf r}'-{\bf R})|^2
\nonumber
\end{eqnarray}
If we notice that the induced part of the nonlocal interaction in the Fermi sea can be written 
as 
\begin{equation}
W^{\textit{ind}}({\bf r},{\bf r}',\omega)=\int d{\bf r}_1\int d{\bf r}_2\ v({\bf r}-{\bf r}_1)\chi({\bf r}_1,{\bf r}_2,\omega)v({\bf r}_2-{\bf r}')
\end{equation}
we find after integration over interaction times  
\begin{eqnarray}
\phi(t)=i \int^{\infty}_{-\infty}\frac{d\omega}{2\pi}\ f(\omega,t)\int d{\bf r}\int d{\bf r}' \hspace{2cm}
\nonumber\\
\label{cuumuluintpot}\\
|u({\bf r}-{\bf R})|^2W^{\textit{ind}}({\bf r},{\bf r}',\omega)|u({\bf r}'-{\bf R})|^2 
\nonumber
\end{eqnarray}
where 
\begin{equation}
f(\omega,t)=\frac{it}{\omega}+\frac{1}{\omega^2}(e^{-i\omega t}-1)
\label{lala}
\end{equation}
is a typical factor characteristic for a spectrum of a structureless hole. 
The first term will give the energy shift of the elastic (no-loss) line, the 
second leads to inelastic structures and provides spectrum normalization.
Because the hole dimension is very small compared to the screening 
length in graphene we can approximate
\[
|u({\bf r}-{\bf R})|^2=\delta({\bf r}-{\bf R})
\]
and the cumulant becomes    
\begin{equation}
\phi({\bf R},t)=i \int^{\infty}_{-\infty}\frac{d\omega}{2\pi}\ f(\omega,t)\ W^{\textit{ind}}({\bf R},{\bf R},\omega).
\label{cumulito}
\end{equation}
For the two-dimensional periodic lattice of graphene, defined by inverse lattice vectors 
${\bf G}$, we can write
\begin{equation}
W^{\textit{ind}}({\bf R},{\bf R},\omega)=\int\frac{d{\bf Q}}{(2\pi)^2}\sum_{\bf G}e^{i{\bf G}\brho}\ W^{\textit{ind}}_{\bf G}({\bf Q},z,z,\omega).
\label{indpotG}
\end{equation}
Calculation of the response function $\chi$ and the induced potential $W^{\textit{ind}}$ in graphene is presented in 
Sec.\ref{resWind}. If we define the spectral function
\begin{equation}
S_{\bf G}({\bf Q},z,\omega)=-\frac{1}{\pi v_{{\bf Q}+{\bf G}}}\Imag\left\{ W^{\textit{ind}}_{\bf G}({\bf Q},z,z,\omega)\right\}
\end{equation}
where $v_{{\bf Q}+{\bf G}}=\frac{2\pi}{|{\bf Q}+{\bf G}|}$ we can write  
\begin{eqnarray}
\phi({\bf R},t)=\hspace{5cm}
\nonumber\\
\label{mala}
\\
\int^{\infty}_{0}d\omega\ f(\omega,t)\ \sum_{\bf G}e^{i{\bf G}\brho}\int\frac{d{\bf Q}}{(2\pi)^2}v_{{\bf Q}+{\bf G}}S_{\bf G}({\bf Q},z,\omega).
\nonumber
\end{eqnarray}
By using (\ref{lala}) and (\ref{mala}) the core hole Green's function can be written as  	 
\begin{equation}
G({\bf R},t)=-i\theta(t)e^{-i(\tilde{E}({\bf R})-i\gamma)t}e^{\tilde{\phi}({\bf R},t)}
\label{greens} 
\end{equation}
where 
\begin{equation}
\tilde{\phi}({\bf R},t)=\int^{\infty}_0\frac{d\omega}{\omega}\ \alpha({\bf R},\omega)[e^{-i\omega t}-1]
\end{equation} 
generates inelastic structures in the spectrum and $\tilde{E}({\bf R})=E_0+\Delta E({\bf R})$ 
where 
\begin{equation}
\Delta E({\bf R})=\int^{\infty}_0d\omega\ \alpha({\bf R},\omega)
\label{zjoksi}
\end{equation}
is the position-dependent core hole energy shift.
Here we also defined the new function, dynamical singularity index (sometimes also called asymmetry parameter in the literature):
\begin{equation}
\alpha({\bf R},\omega)=\frac{1}{\omega}\sum_{\textbf{G}_{\parallel}}e^{i\textbf{G}_{\parallel}\brho}\int\frac{d{\bf Q}}{(2\pi)^2}\ v_{{\bf Q}+\textbf{G}_{\parallel}}
S_{\textbf{G}_{\parallel}}({\bf Q},z,\omega).
\label{alfa}
\end{equation}
Finally the core hole spectrum is given by 
\begin{equation}
A({\bf R},\omega)=\frac{1}{2\pi}\int^{\infty}_{-\infty}dt\ e^{i(\omega-\tilde{E}({\bf R}))t}e^{-\gamma|t|}e^{\tilde{\phi}({\bf R},t)} 
\label{spectrum} 
\end{equation}
which is also normalised 
\[
\int d\omega A({\bf R},\omega)=1
\]
and satisfies the spectral sum rule \cite{Mahknj} 
\begin{equation}
\int d\omega \omega A({\bf R},\omega)=E_0.
\label{saspi}
\end{equation}
\section{Analytic discussion of possible singularities in XPS lineshapes}
\label{sec3}
\begin{figure}[h]
\includegraphics[width=\columnwidth]{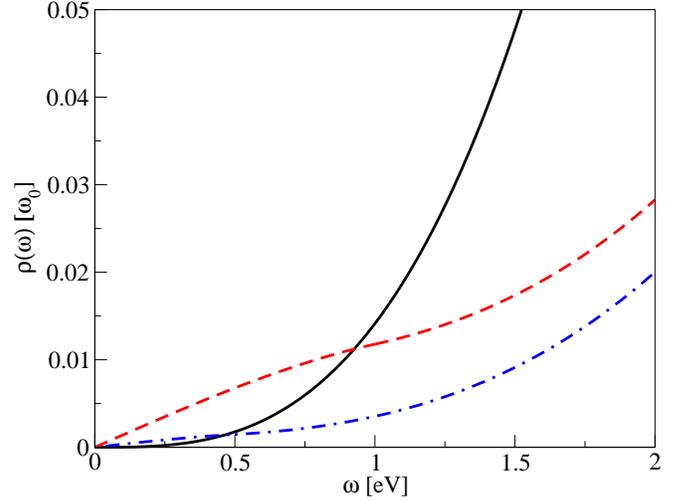}
\caption{Density of electronic excitations $\rho(\omega)$ in pristine graphene (solid line), 
doped graphene ($E_F=0.5$~eV) (blue dotted-dashed line), and doped graphene ($E_F=1$~eV) (red dashed line).}
\label{Fig3a}
\end{figure}

\begin{figure}[h]
\includegraphics[width=\columnwidth]{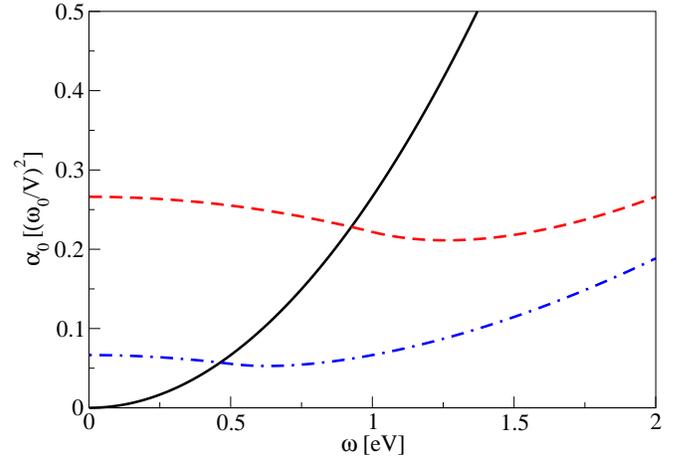}
\caption{Singularity index $\alpha_0(\omega)$ in pristine graphene (solid line), 
doped graphene ($E_F=0.5$~eV) (blue dotted-dashed line), and doped graphene ($E_F=1$~eV) (red dashed line).}
\label{Fig3b}
\end{figure}

In this section we shall estimate the possibility to find singular lineshapes in the localised 
level (i.e. C1s) lineshapes in graphene (or any similar system where electrons form a Dirac cone). 
The relevant quantity is the density of electronic 
excitations $\rho(\omega)$  near the Fermi level. If we assume a fully screened (contact) 
potential V, the dynamical singularity index $\alpha(\omega)$ in (\ref{alfa}) can be reduced 
to    
\begin{equation}
\alpha_0(\omega)=\frac{1}{\omega}|V|^2\rho(\omega)\sim\omega^\beta
\label{jurc}
\end{equation}
which determines the shape of the inelastic contributions to the spectrum  \cite{Mahknj,Bandstr}.
Density of states per unit area of electrons in the Dirac cone, with energies $E_{\bf K}=v |{\bf K}|$, 
for both spin directions is   
\begin{equation}
g(E)=\frac{2}{\omega^2_0}E
\label{densstate}
\end{equation}
where $\omega_0=\sqrt{2\pi}\hbar v/a_0$ and $a_0$ is the Bohr radius. For graphene $\omega_0=47.19$~eV.
The density of excitations (without spin flip) is given by 
\[
\rho(\omega)=\frac{1}{2}\int_{E<E_F,E+\omega>E_F} g(E)g(E+\omega)dE.
\]
In pristine (undoped graphene) $E_F=0$, so using (\ref{densstate}) the density of interband transitions 
becomes 
\begin{equation}
\rho(\omega)=N\int^0_{-\omega} E (E+\omega) dE=\frac{N}{6}\ \omega^3
\end{equation}
where $N=\frac{2}{\omega_0^4}$. Predicted inelastic spectrum for this case ($\beta=2$) goes to zero at the elastic line, which is 
not destroyed by electron scattering \cite{Mahknj}.    
In the case of doped graphene, $E_F>0$, we distinguish two regions. For $\omega<E_F$ only intraband transitions 
are possible, with  the density    
\begin{equation}
\rho(\omega)=N\int^{E_F}_{E_F-\omega} E (E+\omega) dE=N(E_F^2\omega-\frac{1}{6}\omega^3);\ \  \omega<E_F
\label{jukk}
\end{equation}
Linear term in $\omega$ in (\ref{jukk}), i.e. for $\beta=0$, can lead to singular low-energy scattering and therefore singular lineshapes.    
Higher transition energies $\omega>E_F$ are not interesting in this context, but now we can have both intra and 
interband processes: 
\begin{eqnarray}
\rho^{\textit{intra}}(\omega)&=&N\int^{E_F}_{0} E (E+\omega) dE
\nonumber\\
&=&N\left(\frac{1}{3}E_F^3+\frac{1}{2}E_F^2\omega\right),
\nonumber\\
\rho^{\textit{inter}}(\omega)&=&-N\int^{0}_{E_F-\omega} E (E+\omega) dE
\nonumber\\
&=&N\left(\frac{1}{3}E_F^3-\frac{1}{2}E_F^2\omega+\frac{1}{6}\omega^3\right);\ \ \omega>E_F 
\nonumber
\end{eqnarray}
or, taken together    
\begin{equation}
\rho^{\textit{intra}}(\omega)+\rho^{\textit{inter}}(\omega)=N\left(\frac{2}{3}E_F^3+\frac{1}{6}\omega^3\right);\ \ \omega>E_F. 
\end{equation}
Densities of excitations are shown in Fig.\ref{Fig3a}, scaled by the characteristic energy $\omega_0$. 
Singularity indices $\alpha_0$, given by \ref{jurc}, are shown in Fig.\ref{Fig3b} for several $E_F$, but in 
this case we can only illustrate their qualitative behaviour, because $\left|V\right|^2$ is not known in this 
asymptotic approximation. We shall need a full calculation to obtain quantitative results 
for $\alpha(\omega)$, given in Fig.\ref{alpha}.   
But even this analysis indicates that we can expect singular behaviour only in the spectra of doped 
graphene where $\alpha_0(\omega)\sim const.$ for $\omega\rightarrow 0$.
In Sec.\ref{sec5}  we shall verify this conclusion by detailed calculations of $\alpha(\omega)$, 
using graphene wave functions and properly screened interaction 
potential, and show that this asymptotic result is strongly modified.    
\section{Dynamically screened coulomb interaction in  graphene}
\label{sec4}
\subsection{Ground state calculation}
In this section we briefly describe the calculation of the Kohn-Sham (KS) wave functions and energy levels (band structure) in a graphene monolayer which are used to calculate the independent 
electron response function. Schematic representation of a graphene monolayer is shown in Fig.\ref{Fig1}.
For electronic structure calculations we used plane-wave self-consistent field DFT code (PWscf), within the Quantum Espresso (QE) package \cite{QE}, and the 
Perdew-Zunger local density approximation (LDA) for the exchange correlation (xc)-potential \cite{LDA}. An electronic temperature of $k_B T \approx 0.1$ eV was assumed to 
achieve convergence in the calculation of the KS wavefunctions, and all energies were then extrapolated to 0 K. Ground state electronic density was calculated using a 
$12\times12\times1$ Monkhorst-Pack special $K$-point mesh, i.e. by using  $19$ special points in the irreducible Brillouin zone. In the PWscf code we used norm-conserving LDA based pseudopotentials 
for carbon atoms \cite{pseudopotentials}, and we found the energy spectrum to be convergent with a $50$~Ry plane-wave cutoff. Graphene band structure along the high symmetry $\Gamma-K-M-\Gamma$ direction shown in Fig.\ref{Fig2} was calculated along the path with $241$ $k$-points, and it agrees with previous calculations \cite{Bandstr}.  
For the graphene unit cell parameter we used experimental value $a=4.651\ a.u.$ and for the unit cell in $z$ direction (separation between periodically repeated graphene layers) we take $L=5a=23.255\ a.u.$. as is shown in Fig.\ref{Fig1}. For the response function it will be important to choose the right thickness of the electron density, which we have taken to be $2a$, as 
shown in Fig.\ref{Fig1}.       
\subsection{Response function calculation}
\label{resWind}
Independent electron response function matrix for quasi two-dimensional systems can be written as
\begin{equation}
\begin{array}{c}
\chi^{0}_{{\bf G}_\parallel{\bf G}'_\parallel}({\bf Q},\omega,z,z')=
\\
\\
\frac{2}{S}\sum_{{\bf K}\in S.B.Z.}\sum_{n,m}\ \frac{f_n({\bf K})-f_m({\bf K}+{\bf Q})}
{\omega+i\eta+E_n({\bf K})-E_m({\bf K}+{\bf Q})}\times
\\
\\
M_{n{\bf K},m{\bf K}+{\bf Q}}({\bf G}_\parallel,z)\ M^*_{n{\bf K},m{\bf K}+{\bf Q}}({\bf G'}_\parallel,z')
\end{array}
\label{Resfun0}
\end{equation}  
where $S$ is the normalization surface, and in the summation over ${\bf K}$ we have used $101\times101\times1$ $K$-point mesh sampling which corresponds to $10303$ Monkhorst-Pack special $k$-points in the Brillouin zone and $901$ in the irreducible Brillouin zone. Also, $n,m$ summation is carried out over $20$ bands, which proved to be enough for the proper description of the high energy $\pi+\sigma$ plasmon. Damping parameter $\eta$ used in this calculation is $100$~meV. Matrix elements in (\ref{Resfun0}) have the form  
\begin{equation}
M_{n{\bf K},m{\bf K}+{\bf Q}}({\bf G}_\parallel,z)=\left\langle \Phi_{n{\bf K}}\left|e^{-i({\bf Q}+{\bf G}_\parallel)\brho}\right|\Phi_{n{\bf K}+{\bf Q}}\right\rangle_S
\label{Matrel}
\end{equation}
where ${\bf Q}$ and ${\bf G}_\parallel$ are momentum transfer vector and reciprocal lattice vector, respectively, parallel to the $x-y$ plane, integration is performed over the normalization surface $S$. Plane wave expansion of the wave function has the form 
\[
\Phi_{n{\bf K}}(\brho,z)=\frac{1}{\sqrt{V}}e^{i{\bf K}\brho}\ \sum_{\bf G}C_{n{\bf K}}({\bf G})e^{i{\bf G}{\bf r}},
\]
where $V=S*L$ is the normalization volume, ${\bf G}=({\bf G}_\parallel,G_z)$ are $3D$ reciprocal vectors, ${\bf r}=(\brho,z)$ is $3D$ position vector and the coefficients $C_{n{\bf K}}$ are obtained by solving the KS equations. It should be noted that the integration over perpendicular $z$ coordinate in expression (\ref{Matrel}) is still not performed, so the matrix elements are $z$ dependent.       
The RPA response function can be obtained from independent electron response function (\ref{Resfun0}) by solving the Dyson equation
\begin{equation}
\begin{array}{c}
\chi_{\textbf{G}_{\parallel}\textbf{G}_{\parallel}'}(\textbf{Q},\omega,z,z')=
\chi_{\textbf{G}_{\parallel}\textbf{G}_{\parallel}'}^0(\textbf{Q},\omega,z,z')+
\\
\\
\sum_{\textbf{G}_{\parallel 1}}\int^{L/2}_{-L/2}dz_1dz_2\ \chi_{\textbf{G}_{\parallel}\textbf{G}_{\parallel 1}}^0(\textbf{Q},\omega,z,z_1)\times
\\
\\
V(\textbf{Q}+\textbf{G}_{\parallel 1},z_1,z_2)\ 
\chi_{\textbf{G}_{\parallel 1}\textbf{G}_{\parallel}'}(\textbf{Q},\omega,z_2,z')
\end{array}
\label{chiz}
\end{equation}
where $V(\textbf{Q},z,z')=\frac{2\pi}{Q}e^{-Q|z-z'|}$ is a 2D Fourier transform of a 3D bare  coulomb interaction. 
Here it is important to note that $z_1,z_2$ integrations in (\ref{chiz}) are performed within only 
one of the periodically repeated unit cells in the $z$ direction, so the Coulomb interaction with 
other unit cells is excluded. After Fourier expansion of $\chi(z,z')$  
\begin{equation}
\chi_{\textbf{G}_{\parallel}\textbf{G}'_{\parallel}}
(\textbf{Q},\omega,z,z')=
\frac{1}{L}\sum_{G_{z}G_{z}'}
\chi_{\textbf{G}\textbf{G}'}(\textbf{Q},\omega)e^{iG_{z}z-iG'_{z}z'}
\label{nonlocresmat}
\end{equation}
and similarly for $\chi^0(z,z')$, equation (\ref{chiz}) becomes a full matrix equation 
\begin{eqnarray}
\chi_{\textbf{G}\textbf{G}'}(\textbf{Q},\omega)=\chi_{\textbf{G}\textbf{G}'}^0(\textbf{Q},\omega)+\hspace{4cm}
\nonumber\\
\label{chiznew}\\
\sum_{\textbf{G}_{1}\textbf{G}_{2}}\ \chi_{\textbf{G}\textbf{G}_1}^0(\textbf{Q},\omega)
V_{\textbf{G}_{1}\textbf{G}_{2}}(\textbf{Q})
\chi_{\textbf{G}_1\textbf{G}'}(\textbf{Q},\omega)
\nonumber
\end{eqnarray}
where the Coulomb interaction matrix elements have the explicit form 
\begin{equation}
\begin{array}{c}
V_{\textbf{G}_{1}\textbf{G}_{2}}(\textbf{Q})=
\frac{4\pi}{\left|\textbf{Q}+\textbf{G}_1\right|^2}\delta_{\textbf{G}_1\textbf{G}_2}-
p_{G_{z1}}p_{G_{z2}}
\frac{4\pi(1-e^{-\left|\textbf{Q}+\textbf{G}_{\parallel1}\right|L})}
{\left|\textbf{Q}+\textbf{G}_{\parallel1}\right|L}\times
\\
\\
\frac{\left|\textbf{Q}+\textbf{G}_{\parallel1}\right|^2-{G}_{z1}{G}_{z2}}
{(\left|\textbf{Q}+\textbf{G}_{\parallel1}\right|^2+{G}^2_{z1})
(\left|\textbf{Q}+\textbf{G}_{\parallel1}\right|^2+{G}^2_{z2})}
\delta_{\textbf{G}_{\parallel1}\textbf{G}_{\parallel2}}
\end{array}
\label{3Dexm}
\end{equation}
where 
\[
p_{G_z}=
\left\{\begin{array}{ccc}
1;&\ {G_z}=\frac{2k\pi}{L}&
\\
-1;&\ {G_z}=\frac{(2k+1)\pi}{L}&,\ \ k=0,1,2,3,..
\end{array}
\right.
\]
Solution of equation (\ref{chiznew}) has the form 
\begin{equation}
\chi_{\textbf{G}\textbf{G}'}(\textbf{Q},\omega)=
\sum_{\textbf{G}_1}
{\cal E}_{\textbf{G}\textbf{G}_1}^{-1}(\textbf{Q},\omega)
\chi_{\textbf{G}_1\textbf{G}'}^0(\textbf{Q},\omega),
\label{chieps}
\end{equation}
where we have introduced the dielectric matrix 
\begin{equation}
{\cal E}_{\textbf{G}\textbf{G}'}(\textbf{Q},\omega)=
\delta_{\textbf{G}\textbf{G}'}-
\sum_{\textbf{G}_{1}}V_{\textbf{G}\textbf{G}_{1}}(\textbf{Q})\chi^{0}_{\textbf{G}_{1}\textbf{G}'}(\textbf{Q},\omega).
\label{epsm}
\end{equation}
The screened Coulomb interaction then can be 
written as 
\begin{equation}  
W_{\textbf{G}_{\parallel}}(\textbf{Q},\omega,z,z')=\ \delta_{\textbf{G}_{\parallel}0}
v(\textbf{Q},z,z')\ +\ 
W^{\textit{ind}}_{\textbf{G}_{\parallel}}(\textbf{Q},\omega,z,z')
\label{screenint}
\end{equation}
where induced or dynamical part of Coulomb interaction can be written in 
term of matrix elements of response matrix (\ref{chieps})    
\begin{eqnarray}
W^{\textit{ind}}_{\textbf{G}_{\parallel}}(\textbf{Q},\omega,z,z')&\!\!\!=\!\!\!&
\int^{L/2}_{-L/2}dz_1dz_2 
v(\textbf{Q}+\textbf{G}_{\parallel},z,z_1)\times\nonumber\\ 
&&\chi_{\textbf{G}_{\parallel}0}
(\textbf{Q},\omega,z_1,z_2)\ 
v(\textbf{Q},z_2,z').
\label{inddinint}
\end{eqnarray}
After Fourier transformation in real space propagator of induced Coulomb 
intraction becomes 
\begin{eqnarray}
W^{\textit{ind}}(\textbf{r},\textbf{r}',\omega)&\!\!\!=\!\!\!&
\sum_{\textbf{G}_{\parallel}}\int\frac{d{\bf Q}}{(2\pi)^2}\ e^{i(\textbf{G}_{\parallel}+\textbf{Q})\brho}\ 
e^{-i\textbf{Q}\brho'}\times\nonumber\\
&&W^{\textit{ind}}_{\textbf{G}_{\parallel}}(\textbf{Q},\omega,z,z')
\label{Wreal}
\end{eqnarray}
For the RPA response function calculation we have taken the unit cell thickness 
$L=23.255\ a.u.$ which corresponds to five unit cell parameters in the parallel direction. We have neglected crystal local field 
effects in the parallel but not in the perpendicular direction. We have used the energy of $20$ Hartrees as the cut-off for 
Fourier expansion over $G_z$'s which corresponds to $47$ $G_z$ vectors. This cut-off proved to be sufficient to give smooth, 
monotonically decaying tail of induced charge density for $z>a$.
\section{Dynamical singularity index in pristine and doped graphene}
\label{sec5}
In this section we compare the dynamical singularity index $\alpha({\bf R},\omega)$ calculated using properly screened 
potential with the approximations and predictions presented in Sec.\ref{sec3}. 
In the calculation of the spectrum we shall take only the ${\bf G}_{\parallel}=0$ term in (\ref{alfa}), so 
that the parallel coordinate $\brho$ becomes unimportant, and $z=0$. Once we have calculated the function 
$\alpha(\omega)$ we shall also be able to calculate the complete spectrum given by (\ref{spectrum}).     

As expected, numerically calculated functions $\alpha(\omega)$ confirm our predictions of 
their qualitative behaviour in the asymptotic ($\omega\rightarrow 0$) limit, though they differ 
appreciably for higher frequencies where the details of the band structure become important. In the 
asymptotic limit for pristine graphene $\alpha_0(\omega)$ goes to zero, while for the doped 
graphene ($E_F=0.5$ and $1$~eV) it has a finite value.     
\begin{figure}[h]
\includegraphics[width=\columnwidth]{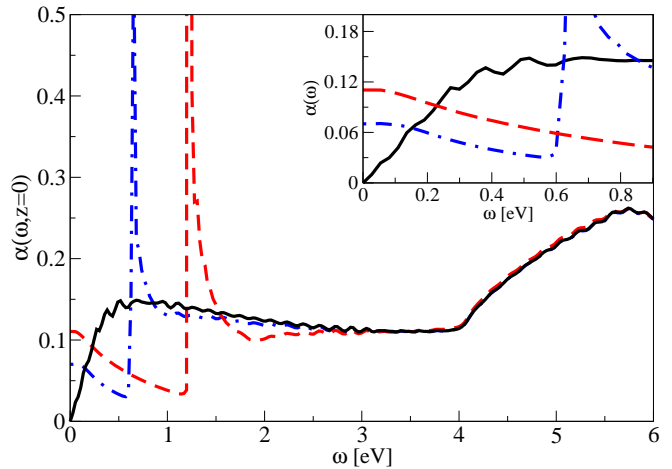}
\caption{Singularity indices in pristine (black solid line), doped $E_F=0.5$~eV (blue dash-dotted line) and doped $E_F=1$~eV 
(red dashed line) graphene.}
\label{alpha}
\end{figure}
In fact, the validity of the asymptotic expression  $\alpha_0(\omega)$ is strongly restricted.
For $E_F=0$ the singularity index  $\alpha(\omega)$ indeed starts from the 
zero, but very soon, on the scale of the core hole linewidth $\gamma$, it increases due 
to  $\pi\rightarrow\pi^*$ interband transitions.       
On the other hand, for $E_F=0.5$ and $1$~eV the $\alpha(\omega)$ starts at finite value, as predicted by the 
asymptotic result (\ref{jurc}), which would indicate singular behaviour, but it immediately 
decreases as the $\pi\rightarrow\pi$ intraband transitions decrease, but a new collective 
mode gives a strong peak at the 2D plasmon frequency \cite{DasSarma}. At even higher energies interband 
$\pi\rightarrow\pi^*$ transitions dominate, and give the same contributions for all graphene 
dopings. We can compare these results with those for a 3D metal \cite{Muller}, where $\alpha(\omega)$ 
is quite constant in a larger $\omega$ region almost up to the appearance of the 
first plasmon peak. Strong variation of $\alpha(\omega)$ in graphene 
could lead to core hole lineshapes showing a combined effect of singular hole relaxation and hole decay processes which all 
occur in the same energy region, as we shall see in Sec.\ref{Allspe}.                    

\section{Calculation of the core-hole spectra} 
\label{sec6}
Once we have calculated the function $\alpha({\bf R},\omega)$ we can calculate complete spectrum given by (\ref{spectrum}), e.g. the core hole energy shift $\Delta E$ 
and the strength of the no-loss line $P_0$.
Let us first analyze the case $\beta=0$, i.e. for $\alpha(\omega)\sim const.$ for $\omega\rightarrow 0$.  
It turns out that expanding the exponent in (\ref{spectrum}) is not a satisfactory procedure because already the first term 
in the expansion 
\begin{equation}
A(\omega)=P_0\delta(\tilde{\omega})+P_0\frac{\alpha(\tilde{\omega})}{\tilde{\omega}}+\ ...
\label{expa}
\end{equation}
where $\tilde{\omega}=\omega-\tilde{E}$, would vanish, and the expansion (\ref{expa}) is therefore meaningless.  
If we approximate  $\alpha(\omega)$ by its asymptotic (constant) value $\alpha=\alpha(0)$, we can calculate the 
spectrum (\ref{spectrum}) analytically to obtain the Doniach-\v Sunji\' c (DS) asymmetric 
lineshape \cite{DSlineshape} 
\begin{equation} \label{SunjDon}
A_{DS}\left(\tilde{\omega}\right) = \frac{1}{\pi}\frac{\Gamma\left(1-\alpha\right)
\cos\left[\frac{\pi\alpha}{2} + \left(1-\alpha\right)\arctan\left(\frac{\tilde{\omega}}{\gamma}\right)\right]}{\left(\tilde{\omega}^{2}+\gamma^{2}\right)^{\left(1-\alpha\right)/2}}
\end{equation}
with the maximum at
\[
\tilde{\omega}_{max}=\gamma\, \cot\frac{\pi}{2-\alpha}.
\]
Incidentally, this maximum is not the shifted elastic line (which is suppressed in this case) 
but corresponds to the inelastic structure due to a large number of soft electron-hole pairs.
This expression is correct only in the low-energy (e.g. $\omega<E_F$) part of the spectrum.  
However, this is not possible for the frequency dependent $\alpha(\omega)$, so we shall instead use a more general 
approach \cite{Minh1,Singindex}. We first notice that the core hole Green's 
function (\ref{greens}) satisfies the equation    
\begin{equation}
\left\{i\frac{\partial}{\partial t}-\tilde{E}+i\gamma\right\}G(t)=\delta(t)+G(t)\int^{\infty}_0d\nu\ \alpha(\nu)e^{-i\nu t}
\label{greenfeq}
\end{equation}
After Fourier transformation we obtain an integral equation for the Green's function
\begin{equation}
G(\tilde{\omega})=G_0(\tilde{\omega})+G(\tilde{\omega})\int^{\infty}_0d\nu\ \alpha(\nu)G(\tilde{\omega}-\nu)
\label{greenfeqinte}
\end{equation}
which can be separated into real and imaginary parts $G_R$ and $G_I$ respectively: 
\begin{eqnarray}
G_{R}\left(\tilde{\omega}\right) = G_{0R}\left(\tilde{\omega}\right)\left[1+J_{R}
\left(\tilde{\omega}\right)\right] - 
G_{0I}\left(\tilde{\omega}\right)J_{I}\left(\tilde{\omega}\right)
\label{Minn1}\\
\nonumber\\
G_{I}\left(\tilde{\omega}\right) = G_{0I}\left(\tilde{\omega}\right)
\left[1+J_{R}\left(\tilde{\omega}\right)\right] + G_{0R}\left(\tilde{\omega}\right)
J_{I}\left(\tilde{\omega}\right)
\label{Minn2}
\end{eqnarray}
where 
\[
J_{R,I}\left(\tilde{\omega}\right) = \int^{\infty}_{0}d\nu \alpha\left(\nu\right)G_{R,I}\left(\tilde{\omega}-\nu\right)
\]
and
\begin{eqnarray} \label{G0}
G_{0R}\left(\tilde{\omega}\right)=\frac{\tilde{\omega}}{\tilde{\omega}^2+\gamma^2} & &
G_{0I}\left(\tilde{\omega}\right)=-\frac{\gamma}{\tilde{\omega}^2+\gamma^2}
\end{eqnarray}
Equations (\ref{Minn1}) and (\ref{Minn2}) can be solved by using the iterative procedure which starts with $G_{0}$ given by (\ref{G0}). 
The core hole spectra can be calculated from
\begin{equation}
A\left(\tilde{\omega}\right) = -\frac{1}{\pi}G_{I}\left(\tilde{\omega}\right)
\label{cuc}
\end{equation}
with $G_{I}$ obtained self-consistently  from (\ref{Minn1}) and (\ref{Minn2}), and for the 
singularity index $\alpha\left(\nu\right)$  calculated from (\ref{alfa}). 
This method enables us to calculate the whole normalised spectrum, but we shall first 
analyze the strength of the no-loss peak, i.e. the 
$\tilde{\omega}=0$ pole contribution to the spectrum in the 
Lorentzian form      
\[
A_{0}\left(\tilde{\omega}\right)=\frac{1}{\pi}\frac{\gamma}{\tilde{\omega}^2+\gamma^2}
\]
From (\ref{Minn2}) one finds 
\[
A\left(\tilde{\omega}\rightarrow 0\right)\rightarrow 
Z\left(\gamma\right)A_{0}\left(\tilde{\omega}\right)
\]
for the decay constant $\gamma$, where $Z\left(\gamma\right)$ is the 
strength of the residuum,
\[
Z\left(\gamma\right)=1+J_{R}\left(\tilde{\omega}=0\right),
\]
or alternatively,  
\[
Z\left(\gamma\right)=\pi\gamma A\left(\tilde{\omega}=0\right).
\]
\section{Discussion}
\label{Allspe}

\begin{figure*}[!t]
\includegraphics[width=\textwidth]{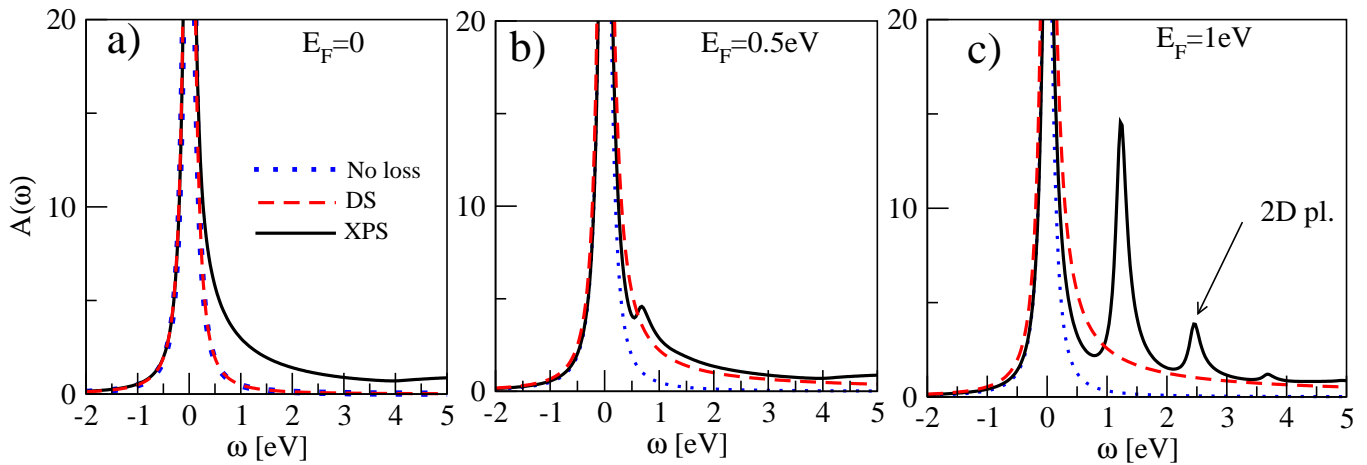}
\caption{a) XPS core hole spectra in pristine graphene (black solid line); DS lineshape (red dashed line); no loss 
line (blue dotted line). b) The same as for a) for doped graphene ($E_F=0.5$~eV). 
c) The same as for a) for doped graphene ($E_F=1$~eV). Hole decay constant $\gamma$ is taken to be $100$~meV.}
\label{Spectra}
\end{figure*}

\begin{figure}[h]
\includegraphics[width=\columnwidth]{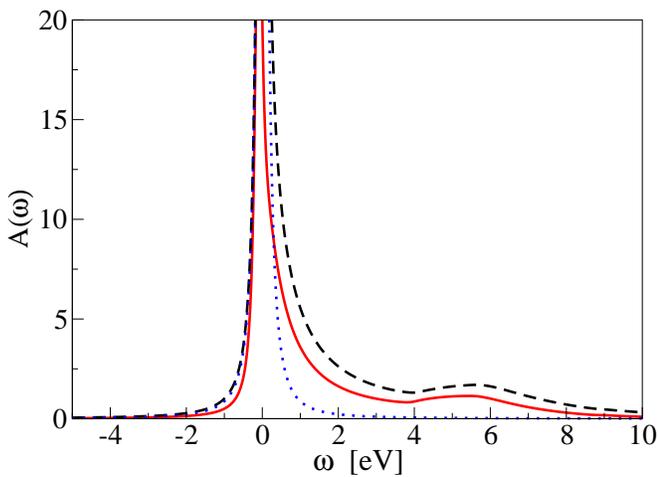}
\caption{Comparison between the no loss line $A_0$ (blue dotted line), first order spectral function $A_1$ (red solid line) and full spectral function 
$A$ (black dashed line).}
\label{Spectra1}
\end{figure}

In Fig.\ref{Spectra} we show core hole spectra calculated using the formalism of Sec.\ref{sec6} 
for pristine graphene ($E_F=0$) and two dopings, $E_F=0.5$ and $1$eV. 
In pristine graphene $\alpha(0)=0$, so DS lineshape $A_{DS}(\omega)$ reduces to the 
Lorentzian $A_0(\omega)$. However, if we depart from this approximation and calculate the spectrum with 
the full $\alpha(\omega)$, we obtain a noticeable low energy tail, which is due to the interband 
$\pi\rightarrow\pi^*$ transitions, as is visible in a rapid increase of $\alpha(\omega)$ in 
Fig.\ref{alpha}. It is easy to mistake it for a singular many-electron tail, i.e. to interpret the 
lineshape in Fig.\ref{Spectra}a as the DS lineshape, e.g. in Fig.\ref{Spectra}b. 
In order to clarify this we have analyzed the spectrum $A(\omega)$ in Fig.\ref{Spectra}a by 
calculating the spectrum $A_1(\omega)$, which includes only first order processes 
(Fig.\ref{Spectra1 }), and we see that they give a dominant contribution to this low energy tail.

For finite doping, in Figs.\ref{Spectra}b,c the DS lineshape given by (\ref{SunjDon}) shows a pronounced 
asymmetry due to $\pi^*\rightarrow\pi^*$ intraband transitions, increasing with doping. 
However, full calculation modifies this asymptotic result as the series of discrete 2D plasmon peaks 
appears in Fig.\ref{Spectra}c at energies $1.2$, $2.4$ and $3.6$eV.

\begin{figure}
\includegraphics[width=\columnwidth]{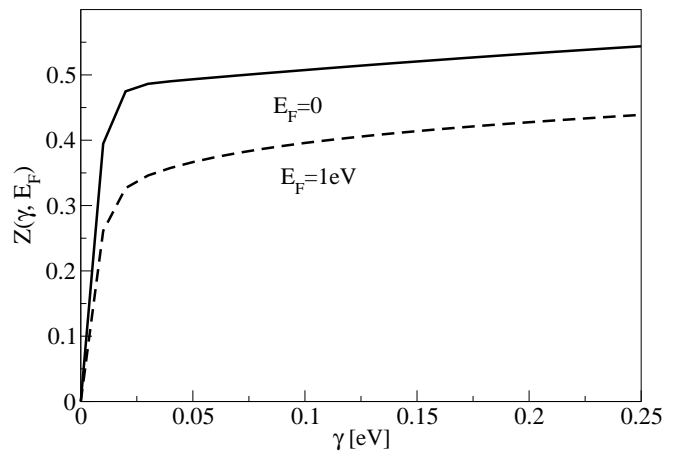}
\caption{The strength of the no-loss line as function of hole decay constant $\gamma$ for pristine graphene $E_F=0$ (solid line) and 
doped graphene $E_F=1$~eV (dashed line).}
\label{Noloss}
\end{figure}

Important information about the strength of many-electron 
excitations can be obtained from the strength of the no-loss line $Z(\gamma,E_F)$, which depends 
on the doping $E_F$ and the hole decay constant $\gamma$, and is shown in Fig.\ref{Noloss}. 
We see that the elastic line is fully destroyed only for extremely small hole decay constants $\gamma$, but 
for realistic values it is substantially reduced, indicating strong inelastic scattering which amounts to $40-50\%$ of the 
total spectral weight.

\begin{figure}
\includegraphics[width=0.8\columnwidth]{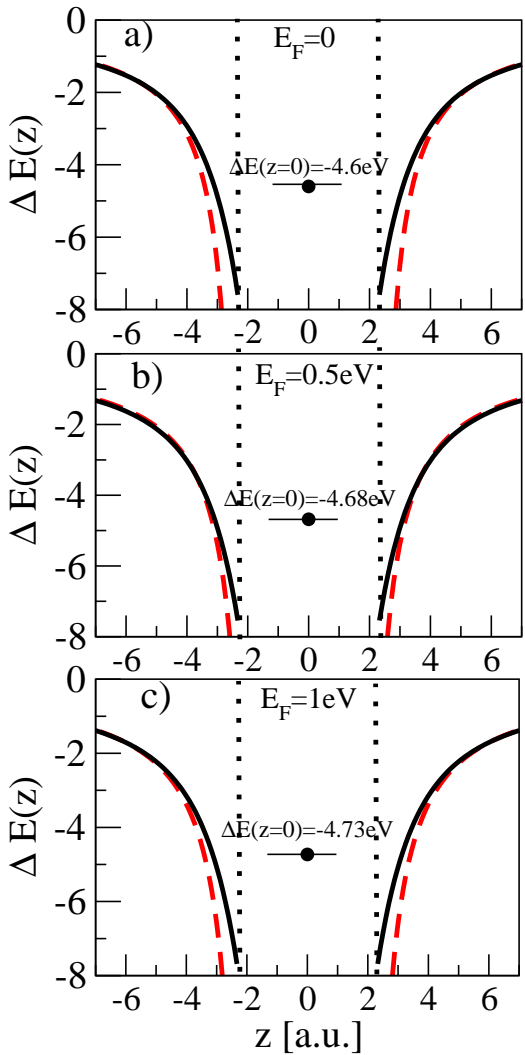}
\caption{Core hole energy shift as function of its z-position (black solid line) in a) pristine graphene, b) doped graphene $E_F=0.5$~eV and c) doped 
graphene $E_F=1$~eV. Corresponding image potential fit (red dashed line) is taken from Ref.\cite{Duncan2}. Dotted vertical lines denote unit supercell 
edges, and black dot represents core hole energy shift in the center of graphene.}
\label{Enshift}
\end{figure}

Another related quantity confirming this conclusion is the ground-state energy shift $\Delta E$, Eq.\ref{zjoksi}, which is connected with the 
total inelastic spectrum by the spectral 
sum rule (\ref{saspi}), so that 
\begin{equation}   
\Delta E= \int d\omega\omega A_{inel}(\omega)
\end{equation}
where $A_{inel}(\omega)=A(\omega)-A_{0}(\omega)$.
Fig.\ref{Enshift} shows core hole energy shift as a function of the core hole position and graphene doping. 
First we observe that the core hole energy shift does not depend on the hole decay constant $\gamma$, as 
can be shown analytically from Eq.\ref{spectrum}. When the core-hole is outside graphene (beyond the graphene electronic density edge) the energy shift follows the 
image potential curves (taken from Ref.\cite{Duncan2}) up to very close distances (4.a.u. from the graphene center) and is mostly due to 
the long-range $\pi\rightarrow\pi^*$ derived plasmon excitations. At shorter distances, as one would expect, the quantum 
mechanical dispersion reduces the polarization shift. Energy shift in the center of the graphene (black dots) 
is smaller than at the unit supercell edges (vertical dotted lines). This behavior is expected, namely the maximum energy shift should 
appear exactly at the centroid of the induced charge which is for graphene at $z_{im}\approx 2.a.u.$ \cite{Duncan2}. This induced charge originates mainly from 
transitions between $\pi$ and $\pi^*$ orbitals. On the other hand $\pi$ orbitals have nodes exactly at the graphene center and that is the 
reason why the energy shift at the graphene center behaves as for the core hole just outside graphene. We notice that the energy shifts in the center of 
graphene, like the effective image plane position $z_{im}$, very weakly depend on graphene doping.  

Let us now compare these theoretical predictions with the experimental observation of singular lineshapes in graphene.
Photoemission spectra involving C1s line in graphene have been reported in a number of papers \cite{a,b,c,d,e,f}, but the primary 
purpose of these measurements was to determine, e.g. structural, chemical or transport properties, growth mechanisms, influence of 
substrate or temperature on these properties, etc. Nevertheless, in several cases measured lineshapes were fitted to the DS asymmetric 
profiles \cite{a,b,e,f}, and even singularity indices were determined \cite{a,b}. 

So, e.g. Gruneis at al \cite{a} fit the measured C1s spectrum to the DS lineshape with $\gamma=216$~meV and 
$\alpha_0=0.1$, which is compatible with the earlier results in Ref.\cite{b} for graphene monolayers deposited on 
various metallic substrates, where DS lineshapes were also used with $\alpha_0$ between $0.1$ and $0.18$. No doping was 
assumed nor discussed in these papers, though it should play an important role in determining the 
asymmetry and possible singular character of core 
hole lineshapes. Also one should consider the influence of the substrate on deposited graphene monolayer. 

The influence of the substrate could be twofold. On one hand it could lead to the charge transfer and doping of the 
graphene monolayer. On the other, if the substrate is metallic, the hole in the graphene 
monolayer can interact with the electrons in the substrate, and also show singular lineshapes, as shown 
in \cite{Singindex}. All these factors should be taken into account when trying 
to determine whether the observed asymmetry is due to singular excitations of 
electron-hole pairs, or to low-order scattering processes.    
In any case, it turns out that the observed asymmetry of the C1s line cannot be directly related 
to the doping in graphene, and more systematic experiments will be needed to 
resolve this interesting issue.  

\section*{Acknowledgments}
The authors are grateful to Donostia International 
Physics Center (DIPC) and Pedro M. Echenique for the hospitality during various stages of this research. 

\end{document}